\documentclass[twocolumn,showpacs,aps,prl]{revtex4}  
\usepackage{epsfig,amssymb,amsmath}
\usepackage[usenames]{color}
\usepackage{ulem}

\newcommand{\vcr}{v_\textrm{cr}}
\begin{document}

\title{Generation of dark-bright soliton trains in superfluid-superfluid counterflow}

\author{C. \surname{Hamner}}
\author{J.J. \surname{Chang}}
\author{P. \surname{Engels}}
\email{engels@wsu.edu} \affiliation{Washington State University, Department of Physics and Astronomy, Pullman,
Washington 99164, USA}
\author{M. A. \surname{Hoefer}}
\affiliation{North Carolina State University, Department of Mathematics, Raleigh, NC 27695, USA}


\begin{abstract}
  The dynamics of two penetrating superfluids exhibit an intriguing
  variety of nonlinear effects.  Using two distinguishable components
  of a Bose-Einstein condensate, we investigate the counterflow of two
  superfluids in a narrow channel. We present the first experimental
  observation of trains of dark-bright solitons generated by the
  counterflow. Our observations are theoretically interpreted by
  three-dimensional numerical simulations for the coupled
  Gross-Pitaevskii (GP) equations and the analysis of a jump in the
  two relatively flowing components' densities.  Counterflow induced
  modulational instability for this miscible system is identified as
  the central process in the dynamics.
\end{abstract}

\maketitle

\par
Nonlinear structures in dilute-gas Bose-Einstein condensates (BECs)
have been the focus of intense research efforts, deepening our
understanding of quantum dynamics and providing intriguing parallels
between atomic physics, condensed matter and optical systems.  For
superfluids that are confined in a narrow channel, one of the most
prominent phenomena of nonlinear behavior is the existence of solitons
in which a tendency to disperse is counterbalanced by the
nonlinearities of the system. In single-component BECs, dark and
bright solitons, forming local density suppressions and local bumps in
the density, resp., have attracted great interest
\cite{Kevrekidis2009}.  In two-component BECs, the dynamics are even
richer as a new degree of freedom, the relative flow between
the two components, is possible.

In this Letter, we investigate novel
dynamics of superfluid-superfluid counterflow, which is in contrast to
the extensively studied counterflow of a superfluid and normal fluid
in liquid helium \cite{Donnelly1991}. Previous theoretical analysis
has demonstrated that spatially uniform, counterflowing superfluids
exhibit modulational instability (MI) when the relative speed exceeds
a critical value \cite{Law2001}. Modulational instability is
characterized by a rapid growth of long wavelength, small amplitude
perturbations to a carrier wave into large amplitude modulations. The
growth is due to the nonlinearity in the system
\cite{Zakharov2009}. Our experiments and analysis reveal that by
carefully tuning the relative speed slightly above the critical value,
we can enhance large amplitude density modulations at the overlap
interface between two nonlinearly coupled BEC components while
mitigating the effects of MI in the slowly varying background regions.
A dark-bright soliton train then results.

In two previous experiments, individual dark-bright solitons were
engineered in two stationary components using a wavefunction
engineering technique \cite{Anderson2001,Becker2008}. In our
experiment we find that trains of dark-bright solitons can occur quite
naturally in superfluid counterflow. This novel method of generating
dark-bright solitons turns out to be robust and repeatable.  In
single-component, attractive BECs the formation of a bright soliton
train from an initial density jump has been predicted
\cite{Kamchatnov2003}.  However, both condensate collapse and the
effects of MI in the density background must be avoided, placing
restrictions on the confinement geometry and diluteness of the
single-component condensate.  In contrast, the properties of
counterflow in miscible, two-component BECs, as we show, enable the
observation of trains consisting of ten or more dark-bright solitons
in BECs with a large number of atoms.  We also note that modulated
soliton trains have been studied extensively in single-component,
modulationally stable repulsive BECs where supersonic flow supports
the generation of dispersive shock waves
\cite{Dutton2001}.
The dark-bright soliton train we study in this work occurs when one of
the system's sound speeds becomes complex so that the standard
definition of supersonic flow does not apply.

Our experiments are conducted with BECs confined in a single-beam optical
dipole trap \cite{EPAPS}. We start with an initially perfectly overlapped mixture of atoms in the
$|F,m_{F}\rangle$ = $|1,1\rangle$ and $|2,2\rangle$ hyperfine states
of $^{87}$Rb, with a total of about 450000 atoms.  The scattering
lengths for the two states used in our
experiment
are estimated to be $a_{11}=100.40$ a.u.~and $a_{22} \approx a_{12} =
98.98$ a.u.~\cite{verhaar_predicting_2009}.  Here $a_{11}$ and
$a_{22}$ denote the single species scattering length for the
$|1,1\rangle$ and $|2,2\rangle$ state, respectively, and $a_{12}$ is
the interspecies scattering length. Mean field theory predicts that a
mixture is miscible if $a_{12}<\sqrt{a_{11}\cdot a_{22}}$
\cite{Timmermans1998,Ao1998,Pu1998}. Therefore our system is predicted
to be weakly miscible. In contrast, previous studies of two-component
binary $^{87}$Rb BECs concentrated mostly on the states which are
immiscible \cite{Hall1998,Mertes2007}, with the notable exception of
Weld et al.~\cite{Weld2009}.

\begin{figure} \leavevmode \epsfxsize=3.375in
  \epsffile{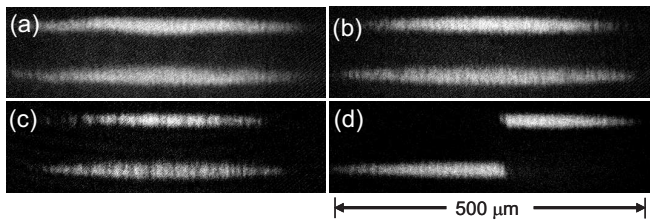} \caption{\label{miscibility} Time evolution
    of an initial perfectly overlapped mixture without (a-c) and with
    (d) an applied axial magnetic gradient. Images taken after
    (a)~100~ms, (b)~1~sec and (c, d)~9~sec of in-trap evolution.}
\end{figure}

When the overlapped mixture is allowed to evolve in the trap, we
observe no phase separation over the experimental timescale of several
seconds. This is in agreement with the predicted miscibility of the
two components and is demonstrated in Fig.~\ref{miscibility}(a-c). The
upper cloud of each image throughout this work shows the atoms in the
$|2,2\rangle$ state at a time 7~ms after a sudden turn-off of the
optical trap (and, where applicable, of any applied magnetic
gradients), while the lower cloud, taken during the same experimental
run, shows the atoms in the $|1,1\rangle$ state after 8~ms of
expansion \cite{EPAPS}. During their in-trap evolution, these clouds are overlapped
in the vertical direction. The dominant effect of the time evolution
in Fig.~\ref{miscibility}(a-c) is a slow decay of the atom number over
time. For single component BECs, we have measured an exponential BEC
lifetime of over 50~sec for the $|1,1\rangle$ state and 14~sec for the
$|2,2\rangle$ state in our dipole trap. Motion induced by changes of
mean field pressure during the decay may be responsible for a small
scale roughness of both components which becomes visible after several
seconds (Fig.~\ref{miscibility}c).

The situation changes when a small magnetic gradient is applied along
the long axis of the trap. Due to Zeeman shifts, the gradient leads to
a force in opposite directions for each component, or equivalently to
a differential shift between the harmonic potentials along the long
axis of the trap. This causes the two components to accelerate in
opposite directions and induces counterflow. In all images where a
magnetic gradient is applied, the gradient is chosen such that the
$|2,2\rangle$ state is pulled to the right and the $|1,1\rangle$ to
the left. An example is shown in Fig.~\ref{miscibility}d where a
gradient leading to a calculated differential trap shift of 60~$\mu$m
was applied for 9~sec, leading to nearly complete demixing of the two
components.

\begin{figure*} \leavevmode \epsfxsize=6.75in
  \epsffile{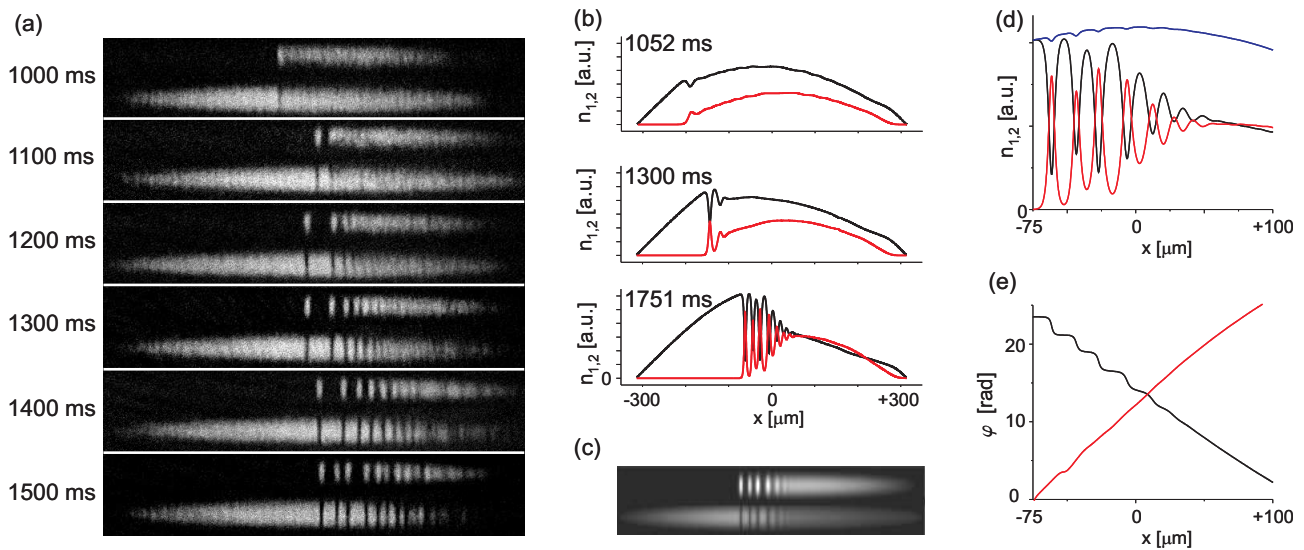} \caption{\label{shocks}
    Formation of dark-bright soliton trains during
    superfluid-superfluid counterflow. A mixture with 70\% of all
    atoms in the $|1,1\rangle$ state (lower cloud in (a) and (c);
    black line in (b),(d),(e)) and 30\% in the $|2,2\rangle$ state
    (upper cloud and red lines, resp.) is created. An axial gradient
    is ramped on over 1~sec and held constant thereafter. (a)
    Experimental images of soliton train formation. Times are measured
    from start of ramp. (b) Integrated cross sections of 3D numerical
    simulations, revealing a gradually steepening overlap interface
    and subsequent soliton formation. (c) Integrated density plot of
    3D numerics at 1751~ms. Field of view 627~$\mu$m $\times$
      17~$\mu$m. (d) Zoomed-in view of soliton train in (b) at
    1751~ms. The blue line shows the total density. (e) Phase
    behavior for region shown in (d).}
\end{figure*}

\begin{figure} \leavevmode \epsfxsize=3.375in
  \epsffile{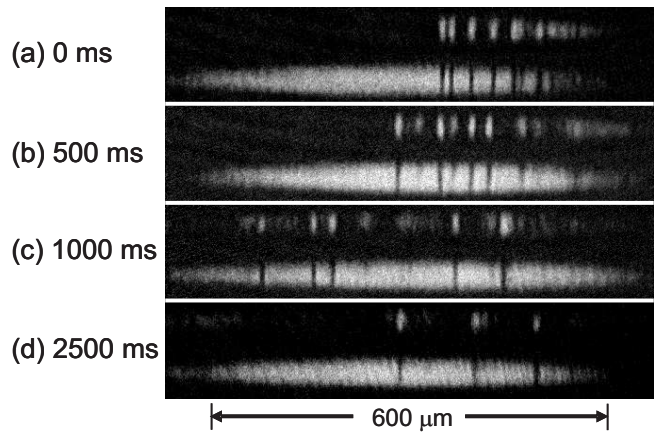} \caption{\label{solitondrift} Motion of
    dark-bright solitons. After creating dark-bright solitons as in
    Fig.~\ref{shocks}(a), the axial magnetic gradient is removed and
    the mixture is allowed to evolve in trap for the evolution times
    given next to each image. The solitons spread out through the
    mixture.}
\end{figure}

In the following we investigate the dynamics induced by small
gradients and show how they can be exploited to create dark-bright
soliton trains. In Fig.~\ref{shocks} an initially overlapped mixture
of 30\% of the atoms in the $|2,2\rangle$ state and 70\% in the
$|1,1\rangle$ state is used. A small magnetic gradient in the axial
direction is linearly ramped on over a timescale of 1~sec, leading to
a calculated trap separation for the two species of only about three
microns. After the end of this ramp, the gradient is held constant.
In the subsequent evolution, individual stripes break off from the
left edge of the $|2,2\rangle$ component, and perfectly aligned dark
notches appear in the $|1,1\rangle$ component
(Fig.~\ref{shocks}(a)). The predominantly uniform widths of the
observed stripes and notches, their long lifetime of several seconds
in the absence of a magnetic gradient, as well as their dynamics resembling individual stable entities (see Fig.~\ref{solitondrift} and below) are strong experimental indications
that the observed features are indeed dark-bright solitons. By
reducing the initial number of atoms in the component forming the bright soliton, we
have also been able to reliably produce one individual dark-bright
soliton and observe its oscillation in trap \cite{Middelkamp2010}, similar to the dynamics
observed in \cite{Becker2008}.

The observed soliton formation is reproduced by three-dimensional (3D)
numerical simulations of the two-component Gross-Pitaevskii (GP)
equations (Fig.~\ref{shocks}(b-e)) \cite{EPAPS}.  Parameters used for
the GP equations are the experimental values.  These values lead to
dynamics that closely match the experiment, as shown in
Fig.~\ref{shocks}(a-c) with a moderate time delay. Our numerical
calculations suggest that the time delay may be due to uncertainties
in the estimated magnetic field gradient induced trap shifts. The
experimentally invoked free expansion directly before imaging the
condensate was not performed in the numerical simulations.

Numerical results for the quantum mechanical phases of the two
wavefunctions describing the components are shown in
Fig.~\ref{shocks}(e). The nearly linear phase behavior on the right
(at $x \gtrsim ~ 50 \mu m$) indicates a smooth counterflow of
the two components. In the soliton region, the phase jumps across the
dark solitons as well as the phase gradients in the bright component
vary slightly, so that the dark-bright solitons are moving relative to
one another which eventually leads to dark-bright soliton
interactions, see \cite{EPAPS}.

The soliton train formation can be qualitatively understood by appealing
to the hydrodynamic formulation of the mean-field, coupled GP
equations in (1+1) dimensions
\begin{align}
  \label{eq:1}
  (\rho_j)_t + (\rho_j u_j)_z &= 0 \\
  \nonumber (u_j)_t + \left (\frac{1}{2} u_j^2 + \rho_j + \sigma_j
    \rho_{3-j} \right )_z &= \frac{1}{4} \left
    [\frac{(\rho_j)_{zz}}{\rho_j} - \frac{(\rho_j)_z^2}{2 \rho_j^2}
  \right ]_z ,
\end{align}
here given in non-dimensional form with $\sigma_j = a_{12}/a_{jj}$,
$\rho_j$ and $u_j$, $j=1,2$ the density and phase gradient (superfluid
velocity) of the $j^\textrm{th}$ component, respectively.  Equation
\eqref{eq:1} models the dynamics of a highly elongated cigar shaped
trap ($\omega_x \sim \omega_y \gg \omega_z$ where $\omega_x$
($\omega_y$) is the transverse trap frequency in the horizontal
(vertical) plane and $\omega_z$ is the axial trap frequency) with
axial confinement neglected \cite{Kevrekidis2009}.  Distance is in
units of the transverse harmonic oscillator length $\sqrt{\hbar/(m
  \omega_x)}$ ($m$ is the particle mass).  Time is in units of
$1/\omega_x$ and the 3D densities are approximated by the harmonic
oscillator ground state via $\rho_j(z,t)
\exp(-x^2-\frac{\omega_y}{\omega_x} y^2)/(2\pi a_{jj} a_0^2)$.

By considering small perturbations proportional to $e^{i(\kappa z -
  \omega t)}$ for uniform counterflow with densities $\rho_j$ and
velocities $u_1 = -v/2$, $u_2 = v/2$, Ref.~\cite{Law2001} demonstrated
modulational instability ($\textrm{Im}\, \omega(\kappa) > 0$) for $v$
larger than a critical velocity $\vcr$ with a maximum growth rate
$\textrm{Im} \, \omega_\textrm{max}$ and associated wavenumber
$\kappa_\textrm{max}$.  We have repeated the calculation and find the
additional result
\begin{equation}
  \label{eq:2}
  \sqrt{\rho_1(1 - \sigma_1 \sigma_2)} \le \vcr \le 2 \sqrt{\rho_1 (1
    - \sqrt{\sigma_1 \sigma_2})} , ~ \rho_1 \ge \rho_2,
\end{equation}
the lower bound being valid for small $\rho_2/\rho_1$ and the upper
bound applicable for $\rho_2 \sim \rho_1$.  The scattering lengths of
the binary system considered here give $0.119 \sqrt{\rho_1} \le \vcr
\le 0.168 \sqrt{\rho_1}$.  Typical densities for the experiments in
Fig.~\ref{shocks} give $\rho_1 + \rho_2 = 4.3$, $\rho_2/\rho_1 = 0.3$
leading to $\vcr = 0.25$ ($\approx 0.22$ mm/s).

\begin{figure} \leavevmode \epsfxsize=3.375in
  \epsffile{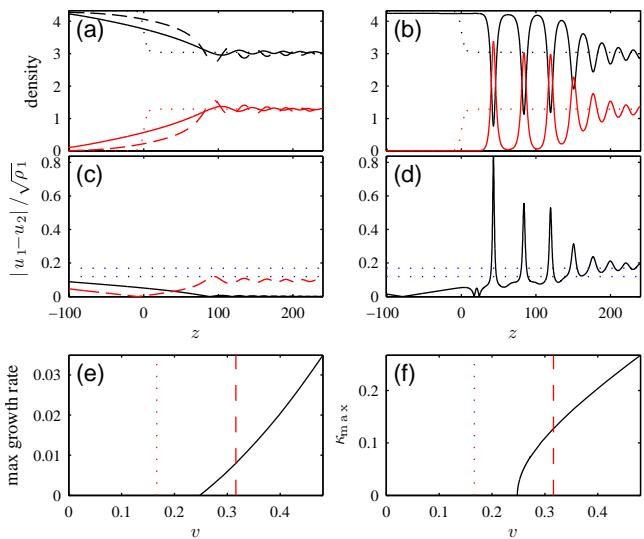} \caption{\label{theory} densities (a,b) and relative speeds
    (c,d) at $t = 462$.
    (e,f): maximum growth rate
    and associated wavenumber of unstable perturbations.
    Vertical lines correspond to subcritical (dotted) and
    supercritical (dashed) cases of (a-d).  See text for further
    details.}
\end{figure}

Figure \ref{shocks} shows that the dark-bright soliton train forms at
the overlap interface of the two components while approximately
maintaining constant total density.  We model this by numerically
solving eq.~\eqref{eq:1} for an initial jump in density that maintains
$\rho_1 + \rho_2 = 4.3$ (dotted curves in \ref{theory}(a,b)) with a
uniform counterflow: $u_1 = -v/2$ and $u_2 = v/2$.  For subcritical
cases $0 \le v < \vcr$, the evolution consists of an expanding
rarefaction wave with weak oscillations on the right edge
(Fig.~\ref{theory}(a); solid line for $v = 0$, dashed line for $v =
0.17$) and corresponding scaled relative speeds $|u_1 -
u_2|/\sqrt{\rho_1}$ (Fig.~\ref{theory}(c) solid, dashed) below
critical (in Fig.~\ref{theory}(c,d), the bounds (\ref{eq:2}) on the
critical velocities are indicated by the dotted lines).  When the
initial relative speed is supercritical, a dark-bright soliton train
forms at the initial jump (Fig.~\ref{theory}(b), $v = 0.32$).  The
relative speed within some regions of the soliton train significantly
exceeds $\vcr$ as shown in Fig.~\ref{theory}(d) suggesting that
counterflow induced MI has the effect of enhancing soliton formation.
Because the initial relative speed $v$ was taken just slightly above
$\vcr$, the maximum growth rate $\textrm{Im} \, \omega_\textrm{max} =
0.0077$ and associated wavenumber $\kappa_\textrm{max} = 0.13$ for
unstable perturbations to the uniform state in the far field are small
(Figs.~\ref{theory}(e,f)).  Therefore, MI in the background
counterflow far from the jump does not develop appreciable magnitude
over the timescale of soliton train formation, in contrast to the
dynamics with $v \gg \vcr$ that we investigate in \cite{Hoefer2010}.

This MI assisted soliton formation technique allows us to create
dark-bright solitons in a well-controlled and repeatable manner, as is
evidenced by the fact that all images of Fig.~\ref{shocks}a form a
very consistent sequence even though they were taken during different
runs of the experiment. In addition to repeatability, future studies
may also require a long lifetime of the solitons.  In single component
BECs, achieving long lifetimes of dark solitons has proven difficult
as they are subject to a transverse instability
\cite{Dutton2001,Anderson2001}. Only recently have dark soliton
lifetimes of up to 2.8~sec been achieved \cite{Becker2008}. It has
been conjectured \cite{Busch2001} and numerically confirmed
\cite{Musslimani2001} that dark-bright solitons are more stable to
transverse perturbations than dark solitons. Experimentally, we indeed
observe long lifetimes of several seconds for the dark-bright
solitons after the magnetic gradient is turned off. The solitons act as individual entities and can move through
the BEC, maintaining their shape for a relatively long time. We
demonstrate this by starting from a situation as in
Fig.~\ref{shocks}(a) at 1.5~sec, where a train of solitons has been
created after the application of an axial magnetic gradient.  When the
gradient is subsequently turned off, the dark-bright solitons move
through the BEC while approximately maintaining their narrow widths
(Fig.~\ref{solitondrift}). The bright and dark part of each individual
soliton remain aligned relative to each other, but any regularity in
the spacing between solitons is lost.  The number of visible solitons decreases over
time, but even after 2.5~sec several solitons are still visible, as in
Fig.~\ref{solitondrift}(d).  Simulations \cite{EPAPS} suggest that
soliton interactions may be the cause of this decay.  In
Fig.~\ref{solitondrift}(c), little diffuse cloudlets of atoms in the
$|2,2\rangle$ state are visible in addition to some solitons, and
corresponding small suppressions of the density in the $|1,1\rangle$
components can be detected. We interpret these features as the decay
products of dark-bright solitons, marking the end of their life cycle.

In conclusion, we have observed dark-bright soliton trains in the
counterflow of two miscible superfluids.  The soliton train is formed
due to relative motion above the critical value for modulational
instability.  By inducing relative speeds slightly above critical, we
can avoid the onset of MI throughout the superfluids over the time
scales of soliton train formation. Together with the long lifetime of
the observed dark-bright solitons, this opens the door to future
experiments with these interesting coherent nonlinear structures.
While the dynamics considered in this Letter are effectively
one-dimensional, a very recent theoretical analysis has shown that
superfluid counterflow in higher dimensions can lead to binary quantum
turbulence, providing another example of the exceptional dynamical
richness of the two-component system \cite{Takeuchi2010}.

\subsection{Acknowledgments}
P.E. acknowledges financial support from NSF and ARO.  M.A.H.
acknowledges financial support from NSF under DMS-0803074, DMS-1008973
and a Faculty Research and Professional Development grant from NCSU.
The authors thank the anonymous referees for beneficial suggestions.

\end{document}